\documentstyle[11pt,epsfig]{article}
\def\g {\gamma}
\def\J {$J/\psi$}
\def\Y {$\Upsilon$}
\def\F {{\cal F}}
\def\g {\gamma}
\def\cc {$c\bar{c}$~}

\def\dc {$D^{*\pm}$~}

\renewcommand{\thefootnote}{\fnsymbol{footnote}}
\def\ap#1#2#3   {{\rm Ann. Phys. (NY)}       #1 (#3) #2}
\def\apj#1#2#3  {{\rm Astrophys. J.}         #1 (#3) #2}
\def\apjl#1#2#3 {{\rm Astrophys. J. Lett.}   #1 (#3) #2}
\def\app#1#2#3  {{\rm Acta. Phys. Pol.}      #1 (#3) #2}
\def\cpc#1#2#3  {{\rm Computer Phys. Comm.}  #1 (#3) #2}
\def\dum#1#2#3  {{~}                         #1 (#3) #2}
\def\epjc#1#2#3 {{\rm Eur. Phys. J. C}       #1 (#3) #2}
\def\err#1#2#3  {{\it Erratum}               #1 (#3) #2}
\def\ib#1#2#3   {{\it ibid.}                 #1 (#3) #2}
\def\jcp#1#2#3  {{\rm J. Comp. Phys.}        #1 (#3) #2}
\def\jmp#1#2#3  {{\rm J. Math. Phys.}        #1 (#3) #2}
\def\ijmp#1#2#3 {{\rm Int. J. Mod. Phys.}    #1 (#3) #2}
\def\jpg#1#2#3  {{\rm J. Phys. G.}           #1 (#3) #2}
\def\mpl#1#2#3  {{\rm Mod. Phys. Lett.}      #1 (#3) #2}
\def\nat#1#2#3  {{\rm Nature (London)}       #1 (#3) #2}
\def\ncim#1#2#3 {{\rm Nuovo Cimento}         #1 (#3) #2}
\def\nim#1#2#3  {{\rm Nucl. Instr. Meth.}    #1 (#3) #2}
\def\np#1#2#3   {{\rm Nucl. Phys.}           #1 (#3) #2}
\def\npb#1#2#3  {{\rm Nucl. Phys. B}         #1 (#3) #2}
\def\pan#1#2#3  {{\rm Phys. At. Nuclei}      #1 (#3) #2}
\def\pl#1#2#3   {{\rm Phys. Lett.}           #1 (#3) #2}
\def\plb#1#2#3  {{\rm Phys. Lett. B}         #1 (#3) #2}
\def\prep#1#2#3 {{\rm Phys. Rep.}            #1 (#3) #2}
\def\prev#1#2#3 {{\rm Phys. Rev.}            #1 (#3) #2}
\def\prc#1#2#3  {{\rm Phys. Rev. C}          #1 (#3) #2}
\def\prd#1#2#3  {{\rm Phys. Rev. D}          #1 (#3) #2}
\def\prev#1#2#3 {{\rm Phys. Rev.}            #1 (#3) #2}
\def\prl#1#2#3  {{\rm Phys. Rev. Lett.}      #1 (#3) #2}
\def\prs#1#2#3  {{\rm Proc. Roy. Soc.}       #1 (#3) #2}
\def\ptp#1#2#3  {{\rm Prog. Theor. Phys.}    #1 (#3) #2}
\def\ps#1#2#3   {{\rm Physica Scripta}       #1 (#3) #2}
\def\rmp#1#2#3  {{\rm Rev. Mod. Phys.}       #1 (#3) #2}
\def\rpp#1#2#3  {{\rm Rep. Prog. Phys.}      #1 (#3) #2}
\def\sjnp#1#2#3 {{\rm Sov. J. Nucl. Phys.}   #1 (#3) #2}
\def\spj#1#2#3  {{\rm Sov. Phys. JETP}       #1 (#3) #2}
\def\spu#1#2#3  {{\rm Sov. Phys.-Usp.}       #1 (#3) #2}
\def\yaf#1#2#3  {{\rm Yad. Fiz.}             #1 (#3) #2}
\def\zp#1#2#3   {{\rm Zeit. Phys.}           #1 (#3) #2}
\def\zpa#1#2#3  {{\rm Zeit. Phys. A}         #1 (#3) #2}
\def\zpc#1#2#3  {{\rm Zeit. Phys. C}         #1 (#3) #2}
\def\et{{\rm et al.}}
\begin{document}
\begin{center}
\begin{large} \begin{bf}
      Heavy quark production \\
in the semihard QCD approach at THERA      
\end{bf} \end{large} \end{center}
~\\~\\
\begin{center}
 S.~P.~Baranov
\footnote{\parbox[t]{10cm}{Electronic address: baranov@sgi.lpi.msk.su}}\\
{\sl P.N.~Lebedev~Physics~Institute,~Leninsky prosp.~53,~Moscow~117924,~Russia}
\\

 N.~P.~Zotov
\footnote{\parbox[t]{10cm}{Electronic address: zotov@theory.npi.msu.su}}\\
{\sl D.V.~Skobeltzyn Institute of Nuclear Pysics,}\\
{\sl M.V.~Lomonosov Moscow State University,~Moscow~119899,~Russia}\\
\end{center}
\vspace*{3cm}~\\
{\bf Abstract}\\
In the framework of the semihard  ($k_T$ factorization) QCD approach,
 we consider the photoproduction of \dc
mesons associated with two hadron jets and the \dc production in DIS
 at THERA conditions with the emphasis on the BFKL and CCFM dynamics of
gluon distributions. In the photoproduction of \dc mesons
 the attention is focused on the variable $x_\g$, which is 
the fraction of the photon momentum contributed to a pair of jets 
with largest $p_T$. We show that our theoretical results are
sensitive to the BFKL type dynamics which may be investigated
at THERA energies. We also discuss possible  effect of \J \, meson spin 
alignement, which is thought to be a vivid manifestation of
gluon off-shellness. 
\\~\\
PACS number(s): 12.38.-t, 13.60.-r, 13.87.Ce\\~\\
{\sl Keywords:} \dc, \J \,and jet production, semihard QCD approach, BFKL and CCFM
unintegrated gluon distribution\\~\\
\newpage
\renewcommand{\thefootnote}{\arabic{footnote}}
\setcounter{footnote}{0}

\noindent\begin{bf}  1.~Introduction\\ \end{bf}

The experimental results on heavy flavour production processes
 obtained  by the H1 \cite{H1} and ZEUS \cite{ZEUS,ZEUS2} collaborations
 at HERA  provide a strong impetus for further theoretical and
experimental studies in a new energy region at THERA conditions. 

In due time,
the experimental data have been compared with next-to-leading order (NLO)
perturbative QCD calculations using the `massive' and `massless' schemes.
The measured cross sections generally lie above the predicted level, and
an agreement between the theoretical and experimental results can only be
achieved using some extreme parameter values. In particular, the production
rates of \dc mesons in the NLO massive scheme \cite{massive} require as low
quark mass as $m_c=1.2$~GeV and as sharp charm fragmentation function as
$\epsilon=0.02$ (in the Peterson parametrization). However, even within this
set of parameters, the shapes of the \dc transverse momentum and rapidity
distributions cannot be said well reproduced.
A good agreement between the massless scheme \cite{massless} and the measured
$p_T(D^*)$ (though not $\eta(D^*)$) spectrum was achieved upon introducing an
additional charm excitation contribution assuming an incredibly large charm
content in the photon structure functions \cite{GorSto}: $c(x)\approx u(x)$. 

The so called $k_T$ factorization, or the semihard approach (SHA)
\cite{GLR}-\cite{CE}, on one hand may give a reasonable solution
 for some of the above problems. On the other hand
the significance of $k_T$ factorization (semihard) approach 
becomes more and more
 commonly recognized. Its applications to a variety of photo-, lepto- and
 hadroproduction processes are widely discussed in the literature 
 \cite{Shab} - \cite{ZotLip}. In many cases a remarkable agreement is
 found between the data and the theoretical calculations regarding the
 photo- \cite{d*gam} and electroproduction \cite{d*dis} of \dc mesons, 
 forward jets \cite{fjets},  
 as well as for specific kinematical correlations observed in the 
 associated \dc{+}jets photoproduction \cite{xgam} at HERA and also 
 the hadroproduction of beauty \cite{Maria, Teryaev}, $\chi_c$ \cite{Teryaev2}
and \J~ \cite{Chao} at Tevatron.
 The theoretical predictions made in ref. \cite{jbfkl} has triggered
 a dedicated experimental analysis \cite{Glad} of the \J~ polarization
 (i.e., spin alignement) at HERA energies.

To some extent, the SHA based on the BFKL \cite{BFKL} gluon dynamics
 includes the relevant
effects of higher order contributions \cite{RSS, LDC}. It has been also
demonstrated in \cite{xgam} that the SHA effectively   
imitates the anomalous coupling of the resolved photon and 
an ad hoc contribution from the resolved photon is no longer needed
at least for the description of the $x_{\g}$ distribution in
\dc photoproduction at HERA energies.

In the present paper we use the semihard QCD approach 
to predict some features of the \dc and \J \, production processes in
the new energy region of THERA collider.\\ 

\noindent\begin{bf}  2.~The semihard QCD approach\\ \end{bf}

 The production of \J ~mesons and open-flavoured \cc pairs is described
 in terms of the photon-gluon fusion mechanism.
 A generalization of the usual parton model to the $k_T$-factorization
 approach implies two essential steps. These are the introduction of
 unintegrated gluon distributions and the modification of the gluon spin
 density matrix in the parton-level matrix elements.

At first we consider the relevant partonic subprocesses.
Let $k_1$, $k_2$, $k_3$ and $p_{\psi}$ be the momenta of the initial state photon,
the initial state gluon, the final state gluon and the final state \J, 
respectively, $\epsilon_1$, $\epsilon_2$, $\epsilon_3$ and $\epsilon_{\psi}$ the
polarization vectors, and $k{=}k_1{+}k_2$. The photon-gluon fusion matrix 
elements then read:
\begin{eqnarray}\label{Mj}
&&\hspace*{-1cm}{\cal M}(\gamma g\to\psi g)=
 tr \{\not\epsilon_1\,(\not p_c - \not k_1 + m_c)
      \not\epsilon_2\,(-\not p_{\bar c} - \not k_3 + m_c)
      \not\epsilon_3\,J(\epsilon_{\psi})\} \nonumber \\ 
&&\times [k_1^2-2(p_c k_1)]^{-1} [k_3^2+2(p_{\bar c}k_3)]^{-1}
    \mbox{\quad + 5 permutations}
\end{eqnarray}
Similarly, for the production of an open-flavoured \cc pair (see fig.~1b):
\begin{eqnarray}\label{Md}
&&\hspace*{-1cm}{\cal M}(\gamma g\to c\bar{c})=\bar{u}(p_c)\{
  \not\epsilon_1\,(\not p_c -\not k_1+m_c)\not\epsilon_2\,
  [k_1^2-2k_1p_c]^{-1} \nonumber \\ &&
 +\not\epsilon_2\,(\not p_c -\not k_2+m_c)\not\epsilon_2\,
  [k_2^2-2k_2p_c]^{-1} \}u(p_{\bar{c}})
\end{eqnarray}
In the expression (\ref{Mj}), the projection operator \cite{BaiBer}
$J(\epsilon_{\psi})=\not{\epsilon_{\psi}}\,(\not{p_c}+m_c)/\sqrt{m_{\psi}}\;$
guarantees the proper spin structure of the \cc state, and the charmed
quarks are assumed to each carry one half of the \J~ momentum,
$p_c$=$p_{\bar c}$=$p_{\psi}/2$, $m_c$=$m_{\psi}/2$. The formation of a meson
from the \cc pair is a nonperturbative process. Within the nonrelativistic
approximation we are using, this probability reduces to a single parameter
related to the meson wave function at the origin $|\Psi(0)|^2$, which is known
for \J~ and \Y~ families from the measured leptonic decay widths.
 
The evaluation of traces in (\ref{Mj})-(\ref{Md}) is straightforward and is doneusing the algebraic manipulation system FORM \cite{FORM}. The complete set of 
matrix elements has been tested for gauge invariance by substituting the gluons
momenta for their polarization vectors.

The multiparticle phase space
$\prod d^3p_i/(2E_i)\,\delta^4(\sum p_{in}-\sum p_{out})$ is parametrized 
in terms of transverse momenta, rapidities and azimuthal angles:
$\frac{d^3p_i}{2E_i}$=$\frac{\pi}{2}dp_{iT}^2dy_i\frac{d\phi_i}{2\pi}$.
Let $\phi_1$ and $\phi_2$ be the azimuthal angles of the initial photon and gluon, 
$\phi_3$, $\phi_{\psi}$, $\phi_c$ and $\phi_{\bar{c}}$ the azimuthal angles of
the partonic subprocess products (i.e., of a \J~ and a gluon, or a charmed quark 
and an antiquark, respectively) and $y_3$, $y_{\psi}$, $y_c$ and $y_{\bar{c}}$ 
their rapidities. Then, the fully differential cross sections read:
\begin{eqnarray}
&& d\sigma(ep\to\psi gX)=
\frac{\alpha_s^2\alpha e_c^2\,|\Psi(0)|^2}{4\,x_2\,s^2}\,
\frac{1}{4}\sum_{\mbox{{\tiny spins}}}\,
\frac{1}{8}\sum_{\mbox{{\tiny colours}}}
|{\cal M}(\g g\to\psi g)|^2  \nonumber \\ && \times
 \F_g(x_2,k_{2T}^2,\mu^2)\,\, \label{lipsj}
dk_{1T}^2\, dk_{2T}^2\, dp_{\psi T}^2\, dy_3\,dy_{\psi}\,
\frac{d\phi_1}{2\pi}\,\frac{d\phi_2}{2\pi}\,\frac{d\phi_{\psi}}{2\pi},\\
&& d\sigma(ep\to c\bar{c}X)=
\frac{\alpha_s\alpha e_c^2}{16\pi\,x_2\,s^2}\,
\frac{1}{4}\sum_{\mbox{{\tiny spins}}}\,
\frac{1}{8}\sum_{\mbox{{\tiny colours}}}
|{\cal M}(\g g\to c\bar{c})|^2  \nonumber \\ && \times
 \F_g(x_2,k_{2T}^2,\mu^2)\,\, \label{lipsd}
dk_{1T}^2\, dk_{2T}^2\, dp_{cT}^2\, dy_c\,dy_{\bar{c}}\,
\frac{d\phi_1}{2\pi}\,\frac{d\phi_2}{2\pi}\,\frac{d\phi_c}{2\pi}.
\end{eqnarray}
The phase space physical boundary is determined by the inequality
\begin{equation}
G(\hat{s}, \hat{t}, k_4^2, k_1^2, k_2^2, k_3^2) \le 0,
\end{equation}
where $k_3$ and $k_4$ denote the partonic subprocess final state momenta,
$\hat{s}=(k_1+k_2)^2$, $\;\hat{t}=(k_1-k_3)^2$, and $G$ is the standard
kinematical function \cite{BycKaj}.

When calculating the spin average of the matrix element squared,
we substitute the full lepton tensor for the photon polarization matrix:
\begin{equation} \label{epsgam}
\overline{\epsilon_1^{\mu}\epsilon_1^{*\nu}}=
4\pi\alpha [8p_e^{\mu}p_e^{\nu}-4(p_ek_1)g^{\mu\nu}]/(k_1^2)^2
\end{equation}
(including also the photon propagator factor and photon-lepton coupling).
For the off-shell incoming gluon we take \cite{GLR}
\begin{equation} \label{epsglu}
\overline{\epsilon_2^{\mu}\epsilon_2^{*\nu}}=
k_{2T}^{\mu} k_{2T}^{\nu}/|k_{2T}|^2.
\end{equation}
This formula converges to the usual
$\sum\epsilon^{\mu}\epsilon^{*\nu}=-g^{\mu\nu}$ when $k_{T}\to 0$.
The final state gluon in (\ref{Mj}) is assumed on-shell,
$\sum{\epsilon_3^{\mu}\epsilon_3^{*\nu}}=-g^{\mu\nu}$.
The \J~ polarization vector $\epsilon_{\psi}$ is defined as an explicit
four-vector. In the frame with z-axis along the \J~ momentum,
$p_{\psi}=(0,\,0,\,|p_{\psi}|,\,E_{\psi})$,
it reads for different helicity states:
\begin{equation}
\epsilon_{\psi(h=\pm 1)} = (1,\,\pm i,\, 0,\, 0)/\sqrt{2} ,\qquad
\epsilon_{\psi(h=0)} = (0,\,0,\, E_{\psi},\,|p_{\psi}|)/m_{\psi}.
\end{equation}

The initial photon and gluon momentum fractions $x_1$ and $x_2$ are calculated
from the energy-momentum conservation laws in the light cone projections:
\begin{eqnarray}
 (k_1+k_2)_{E+p_{||}}=
 x_1\sqrt{s} &=& m_{3T}\exp(y_3)\;\; + \; |m_{4T}|\exp(y_4),
 \nonumber \\[-2mm] && \\[-2mm]
 (k_1+k_2)_{E-p_{||}}=
 x_2\sqrt{s} &=& m_{3T}\exp(-y_3) + |m_{4T}|\exp(-y_4). \nonumber
\end{eqnarray}
The multidimensional integration in (\ref{lipsj}), (\ref{lipsd}) has been
performed by means of Monte-Carlo technique, using the routine VEGAS
\cite{VEGAS}.

Another important ingredient of the semihard approach is the so called
unintegrated gluon distribution ${\cal F}(x,k_T^2,\mu^2)$, which determines
the probability to find a gluon carrying the longitudinal momentum fraction
$x$ and transverse momentum $k_T$.
In calculations we use two different sets of unintegrated gluon
distributions. One of them  is based on the approach of ref.\cite{Bluem}
 and is costructed as a leading-order perturbative solution of the BFKL
 equations. Technically, the unintegrated gluon density
 $\F_g(x,k_{T}^2,\mu^2)$ is calculated as a convolution of the ordinary
 gluon density $G(x,\mu^2)$ with universal weight factors:
\begin{equation} \label{conv}
 {\cal F}_g(x,k_{T}^2,\mu^2) = \int_x^1
 {\cal G}(\eta,k_{T}^2,\mu^2)\,
 \frac{x}{\eta}\,G(\frac{x}{\eta},\mu^2)\,d\eta,
\end{equation}
\begin{equation} \label{J0}
 {\cal G}(\eta,k_{T}^2,\mu^2)=\frac{\bar{\alpha}_s}{\eta\,k_{T}^2}\,
 J_0(2\sqrt{\bar{\alpha}_s\ln(1/\eta)\ln(\mu^2/k_{T}^2)}),
 \qquad k_{T}^2<\mu^2,
\end{equation}
\begin{equation}\label{I0}
 {\cal G}(\eta,k_{T}^2,\mu^2)=\frac{\bar{\alpha}_s}{\eta\,k_{T}^2}\,
 I_0(2\sqrt{\bar{\alpha}_s\ln(1/\eta)\ln(k_{T}^2/\mu^2)}),
 \qquad k_{T}^2>\mu^2,
\end{equation}
where $J_0$ and $I_0$ stand for Bessel functions (of real and imaginary
arguments, respectively), and $\bar{\alpha}_s=\alpha_s/3\pi$. The LO GRV
set \cite{GRV95} was used here for the boundary conditions.
Another set of unintegrated gluon densities
 was extracted from a numerical  
simulution of the CCFM equations \cite{CCFM} and then tabulated
  in the form of a FORTRAN code \cite{fjets}. Finally, the charm quarks
 were converted into \dc
 mesons using the
Peterson fragmentation function \cite{Peterson}.

In the paper \cite{d*gam} we used the standard GRV parametrization
\cite{GRV95} for the collinear gluon density, from which the unintegrated
gluon distribution was developed according to eqs. (\ref{conv})-(\ref{I0}).
Some other essential parameters were chosen as follows: the charm quark mass
$m_c=1.5$~GeV, the Peterson fragmentation parameter $\epsilon=0.06$, the
overall $c\to D^*$ fragmentation probability 0.26.
The Pomeron intercept $\Delta$ was regarded as free parameter, and then the
value $\Delta=0.35$ has been extracted from a fit to the experimental
$p_T(D^*)$ spectrum measured by the ZEUS collaboration \cite{ZEUS}.
Close estimates for $\Delta$ have also been obtained by many other authors,
see, e.g. \cite{Delta, L3}.
Since the agreement with the data achieved within this set of parameters was
really good \cite{d*gam,d*dis}, we continue using it in the present calculations.
\par
\begin{figure*}[ht]
\begin{center}
\includegraphics[width=0.45\linewidth]{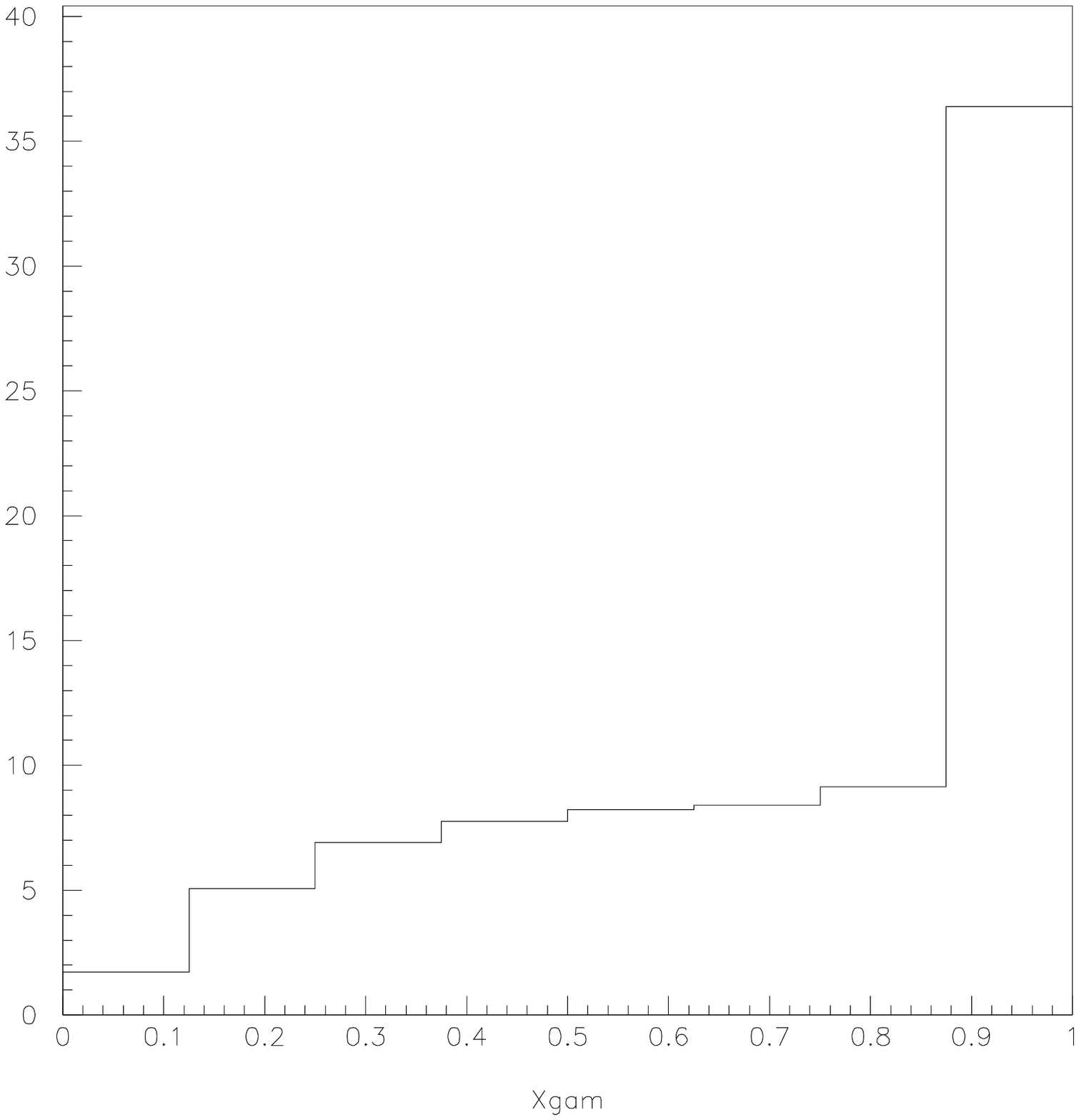}
\includegraphics[width=0.45\linewidth]{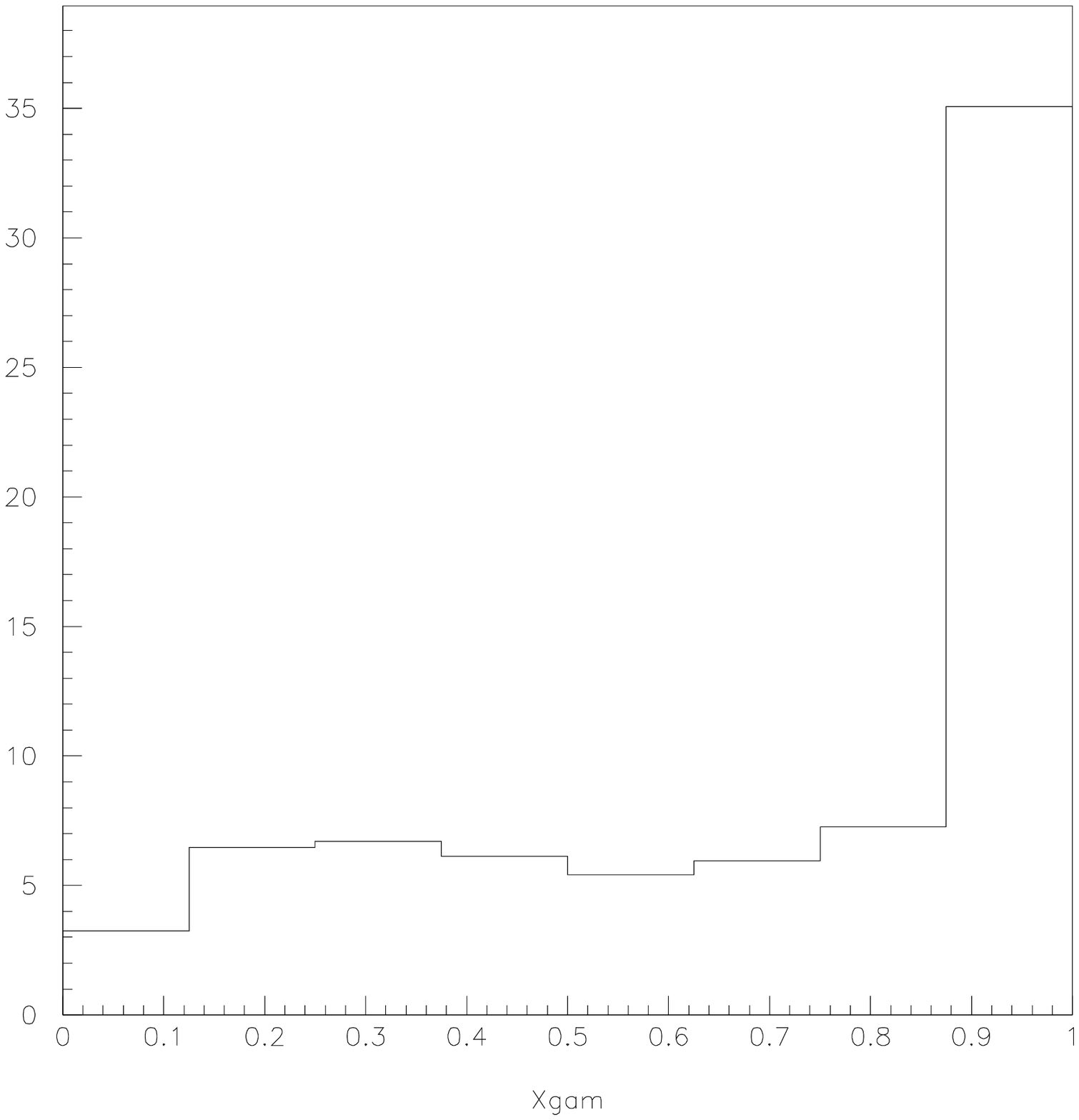}
\end{center}
\vspace*{-5mm}
\caption{{\it The differential cross section $d \sigma /dx_\gamma$ for 
$Q^2<1$ GeV$^2$ with BFKL and  CCFM unintegrated gluon distributions
at THERA.
}}
 \label{xgamcc}
\end{figure*}
\newpage
\noindent\begin{bf}  3.~Numerical results\\ \end{bf}

\noindent\begin{bf} 3.1~ \dc and dijet associated photoproduction
at THERA\\ \end{bf}

The ZEUS collaboration
has measured the associated charm and dijet production \cite{ZEUS} as
a further test of the underlying parton dynamics. 
In these measurements, the quantity of interest is the fraction of the photon
momentum contributing to the production of two jets with highest $E_T$,
which is experimentally defined as
\begin{equation}
x_{\gamma}=[E_{1T}\exp(-\eta_1)+E_{2T}\exp(-\eta_2)]/(2E_e\,y)
\label{xgam} \end{equation}
with $E_{iT}$ and $\eta_i$ being the transverse energy and rapidity of these
hardest jets. 

In the ref. \cite{d*dis} the theoretical calculations have been made
within the SHA with different unintegrated gluon
distributions at HERA energies.
In Fig.~1 we present the results of similar theoretical calculations made within
the semihard approach with BFKL and CCFM unintegrated gluon
distributions at THERA energies.
The simulation procedure consists in generating a photon-gluon fusion event
using the off-shell matrix elements and the unintegrated gluon distribution
functions described in Section 2.

The basic $2\to 2$ partonic subprocess gives rise to two high-energy quarks,
which can further evolve into hadron jets. Actually, as the matter of some
reasonable approximation, the calculations were restricted to parton level,
and so the produced quarks (with their known kinematical parameters) were
taken to play the role of the final jets: $E_T(jet_{1,2})=E_T(q,\bar{q})$.

The two quarks are accompanied by a number of gluons radiated in the course
of the gluon evolution. It has been mentioned already that, on the average,
the gluon transverse momentum decreases from the hard interaction block towards
the proton. As an approximation, we assume that the gluon  closest to the
quark block with its momentum  $k'$ 
compensates the whole transverse momentum of the virtual gluon
participating in the hard interaction: $\vec{k'}_T\simeq -\vec{k}_T$, while
all the other emitted gluons are collected together in the `proton remnant',
which is assumed to carry only a negligible transverse momentum compared to
$\vec{k'}_T$. This gluon gives rise to a third hadron jet with $E_T=|\vec{k'}_T|$.

From among the three hadron jets represented by the quark, antiquark and gluon
we choose the two carrying the largest transverse energies, and then calculate
the quantity $x_\g$ according to its definition given by equ. (\ref{xgam})
\footnote{The full FORTRAN code is available from the authors on request.}.

In a significant fraction of events, the gluon radiated from the BFKL cascade
appears to be harder than one or even both of the quarks produced in hard
parton interaction \cite{xgam}.
In fact, the specified events are responsible for the wide plateau seen in the
$x_\g$ distribution in Fig.~1.\\
\begin{figure*}[ht]
\begin{center}
\includegraphics[width=0.31\linewidth]{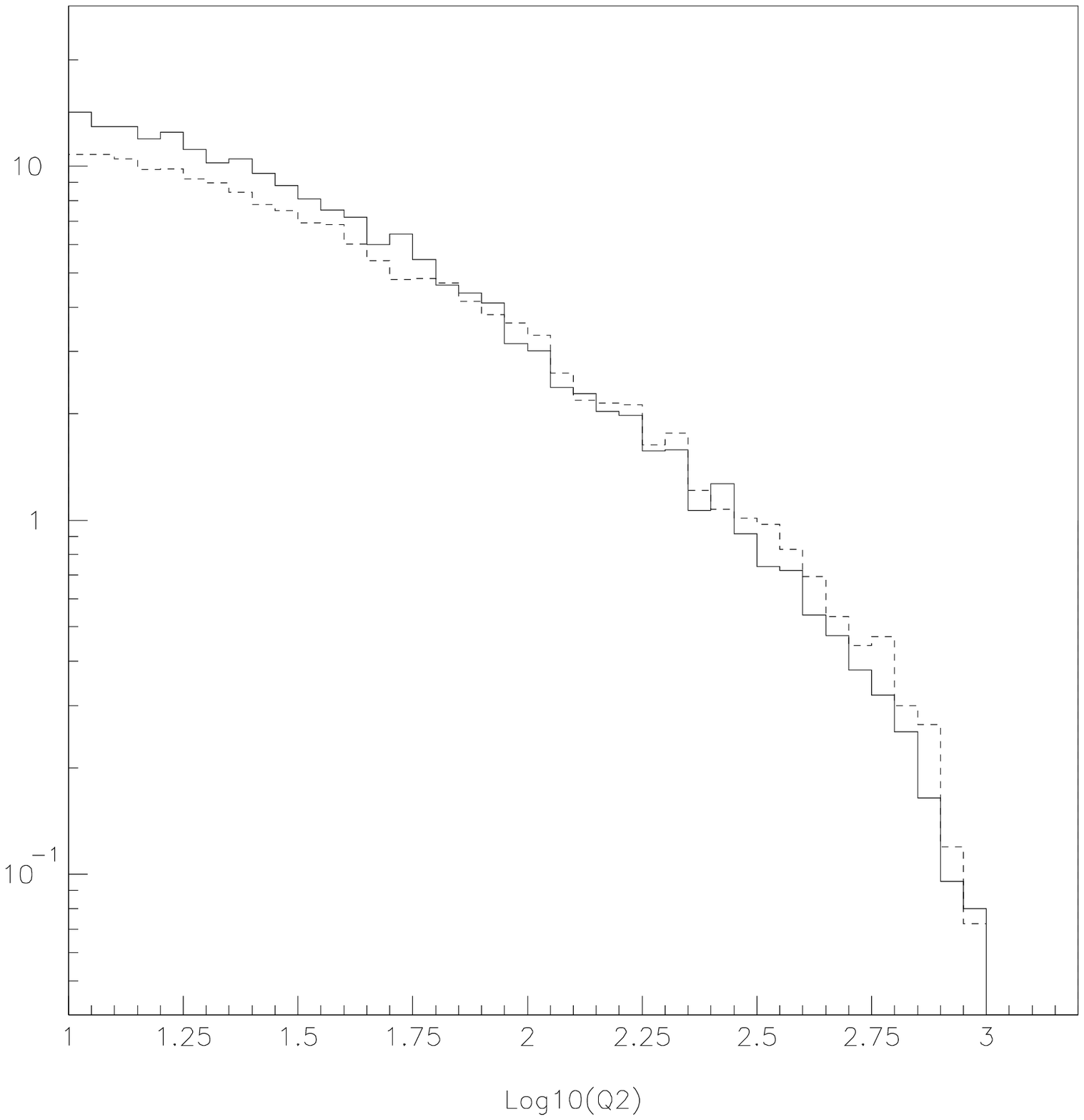}
\includegraphics[width=0.31\linewidth]{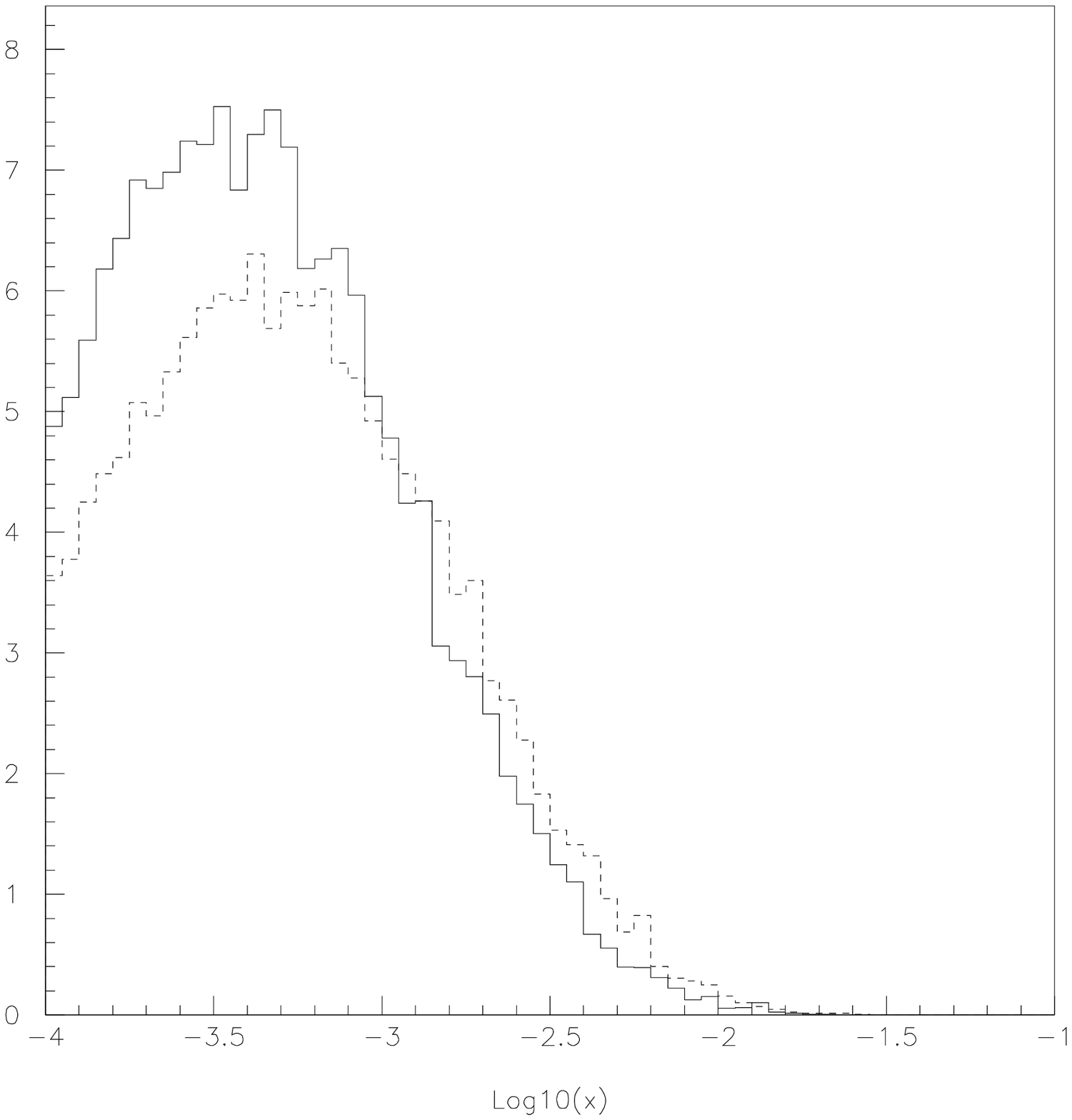}
\includegraphics[width=0.31\linewidth]{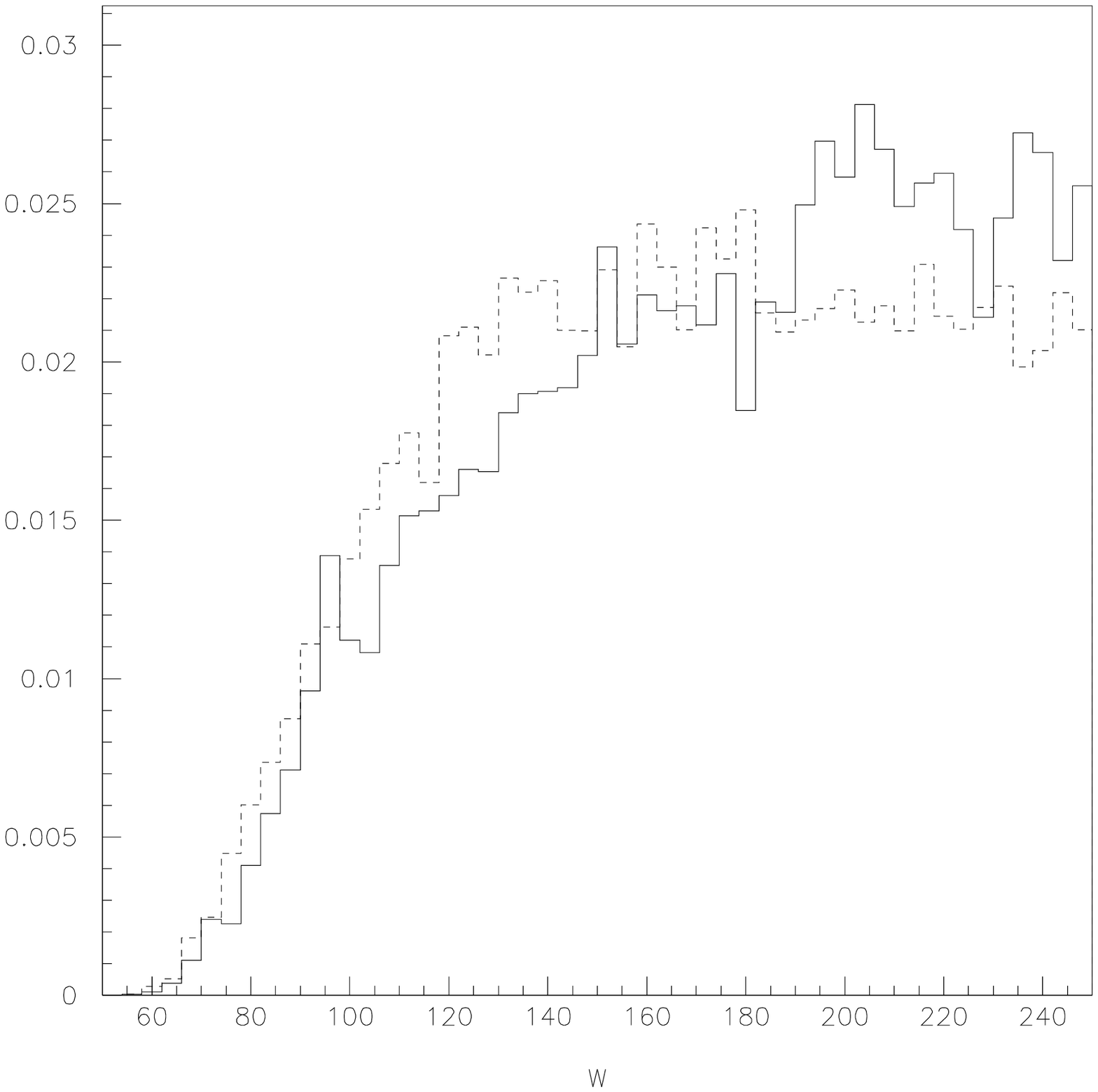}
\includegraphics[width=0.31\linewidth]{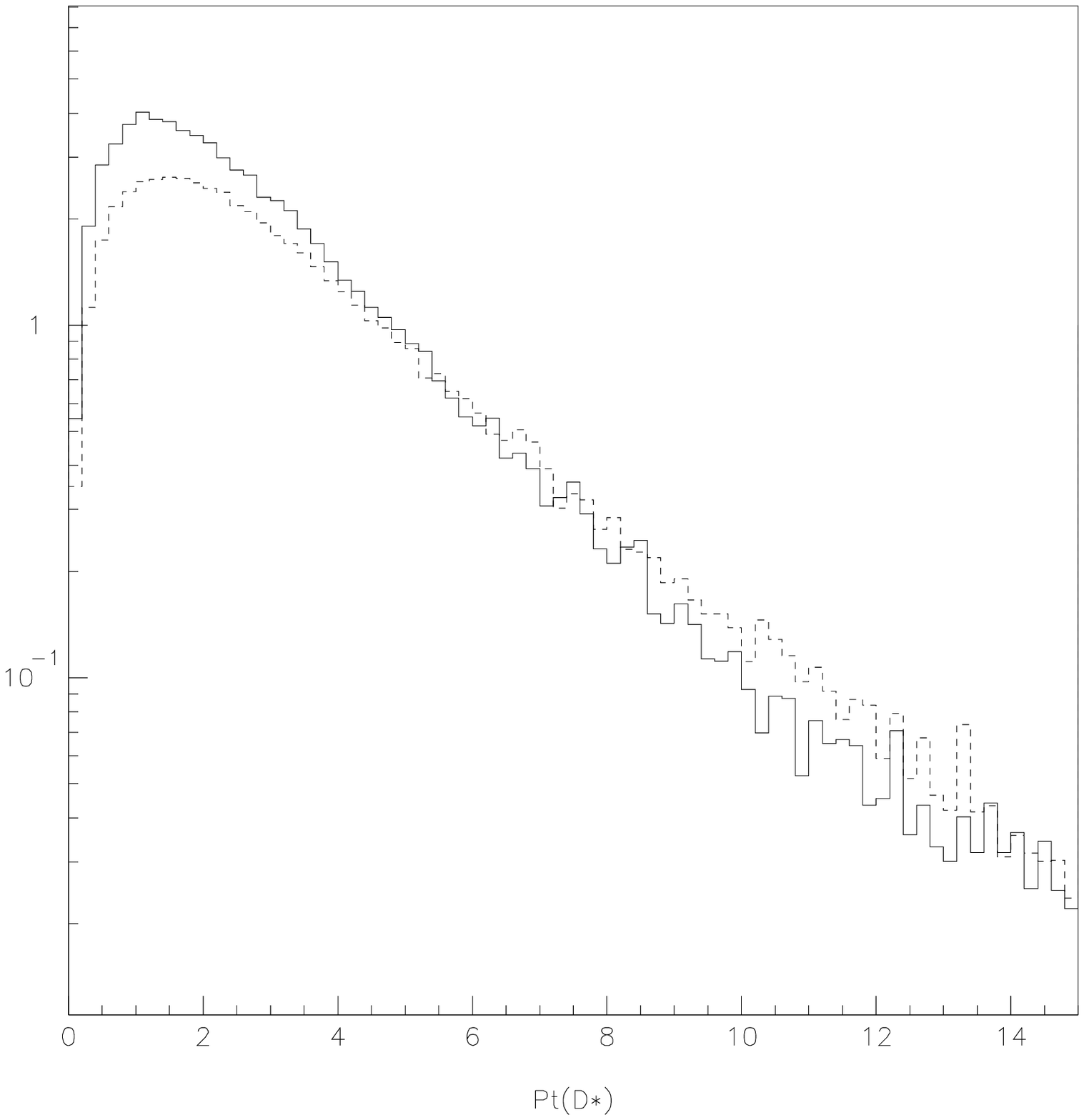}
\includegraphics[width=0.31\linewidth]{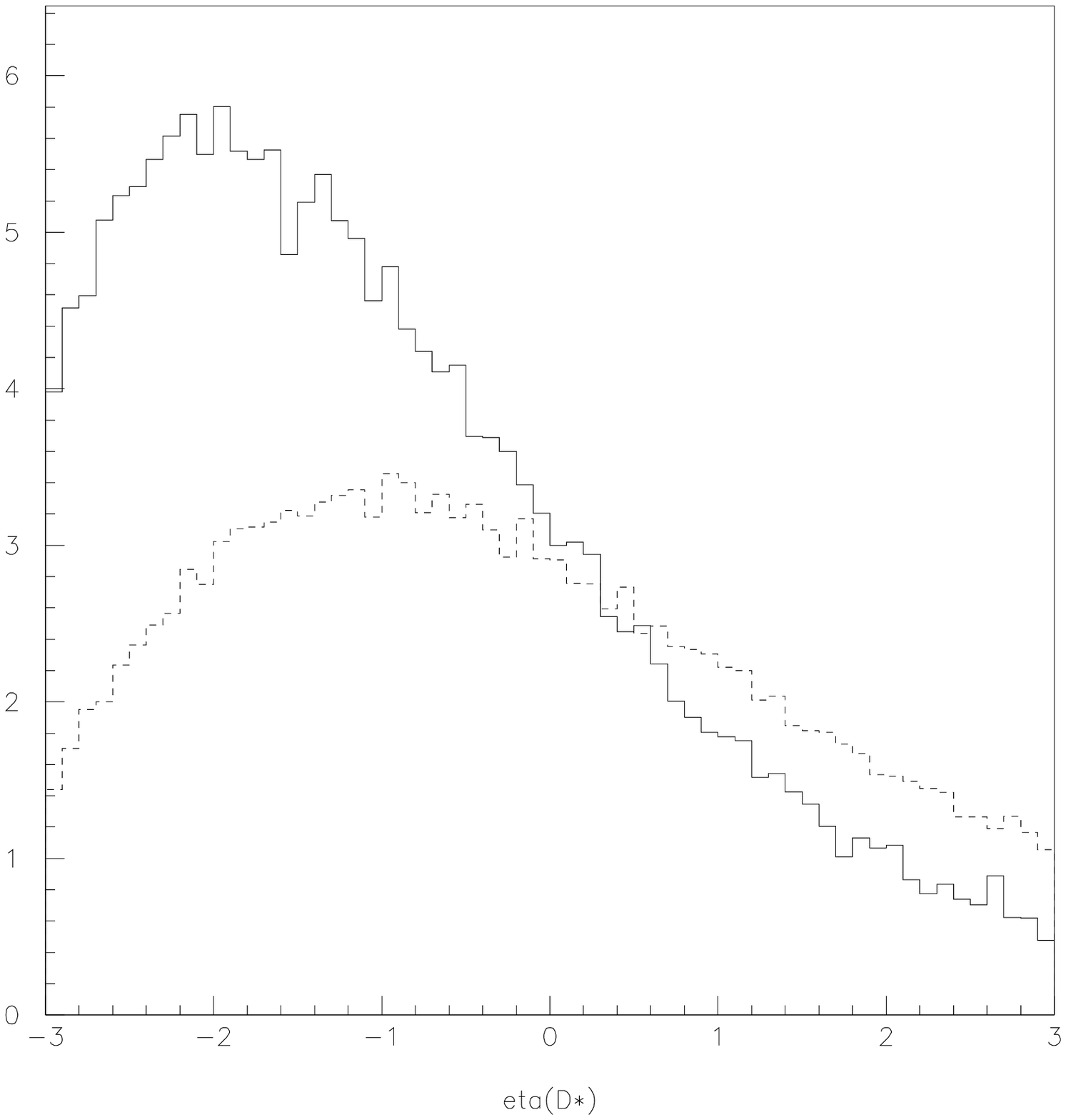}
\includegraphics[width=0.31\linewidth]{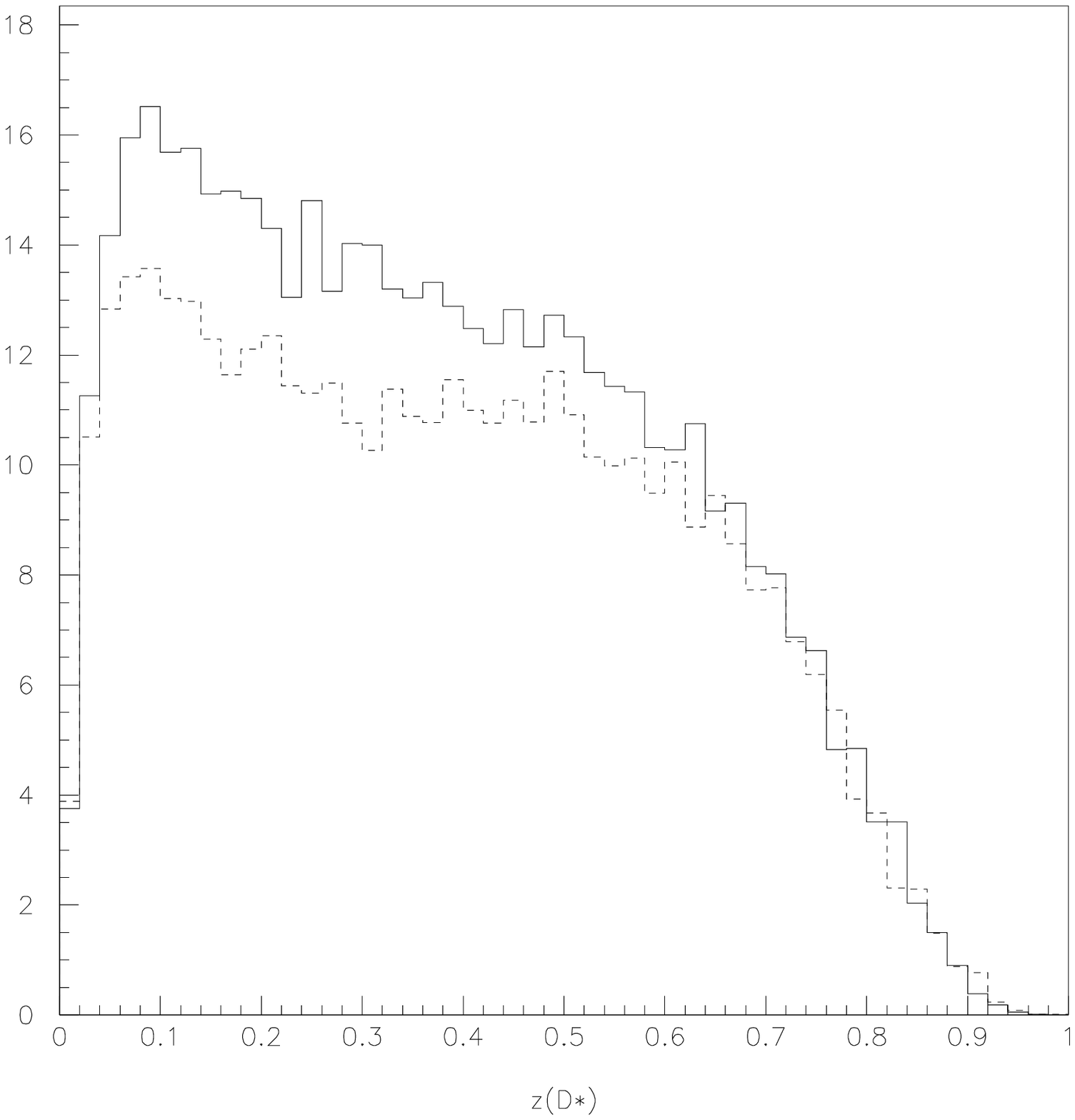}
\end{center}
 \vspace*{-5mm}
\caption{{\it Differential cross sections for deep inelastic $D^{*\pm}$ 
production with BFKL and CCFM unintegrated gluon distributions
at THERA  as functions of:
$(a)\;\log_{10}Q^2$,
$(b)\;\log_{10}x$,
$(c)\;W$,~
$(d)\;p_T(D^*)$,
$(e)\;\eta(D^*)$ and
$(f)\;z(D^*)$.}
 \label{fig:fig1}}
\end{figure*}

\noindent {\bf 3.2. Deep inelastic \dc production at THERA}\\

The process of deep inelastic \dc production at HERA  is truly semihard 
because of the presence of two large scales:
the virtuality of the exchanged photon ($Q^2$) and the charm mass ($m_c^2$),
both being much larger than $\Lambda_{QCD}$ but much smaller than $s$.
Therefore,  experimental data concerning the \dc production in DIS
at THERA provide  a strong impetus for further theoretical studies
of this process.
 
 In Fig.~2 the theoretical predictions on the differential cross sections of
deep inelastic \dc production are shown for the THERA 
kinematical
region: $1\,<\, Q^2\, <\, 1000$ GeV$^2$, \, $1.5\, <\, p_T(D^{*\pm})\, <\, 15$
GeV and $|\eta (D^{*\pm})|\, < \, 1.5$. Different curves in Fig.~2
correspond  to the BFKL and CCFM unintegrated gluon distributions.
At HERA energies the SHA calculations with BFKL unintegrated gluon
 distribution have shown \cite{d*dis} some 
shift to negative $\eta(D^*)$ with respect to the ZEUS data.
  When we have used the CCFM unintegrated gluon density 
from MC generator CASCADE \cite{fjets}
with JETSET based fragmentation function~\cite{JETSET}
implemented in, we obtain  
good agreement between our theoretical results and the ZEUS experimental
data \cite{ZEUS2} also for $d\sigma/d\eta(D^*)$ \cite{d*dis}. \\  

\noindent {\bf 3.3. \J \, photoproduction at THERA}\\
\begin{figure*}[ht]
\begin{center}
\includegraphics[width=0.31\linewidth]{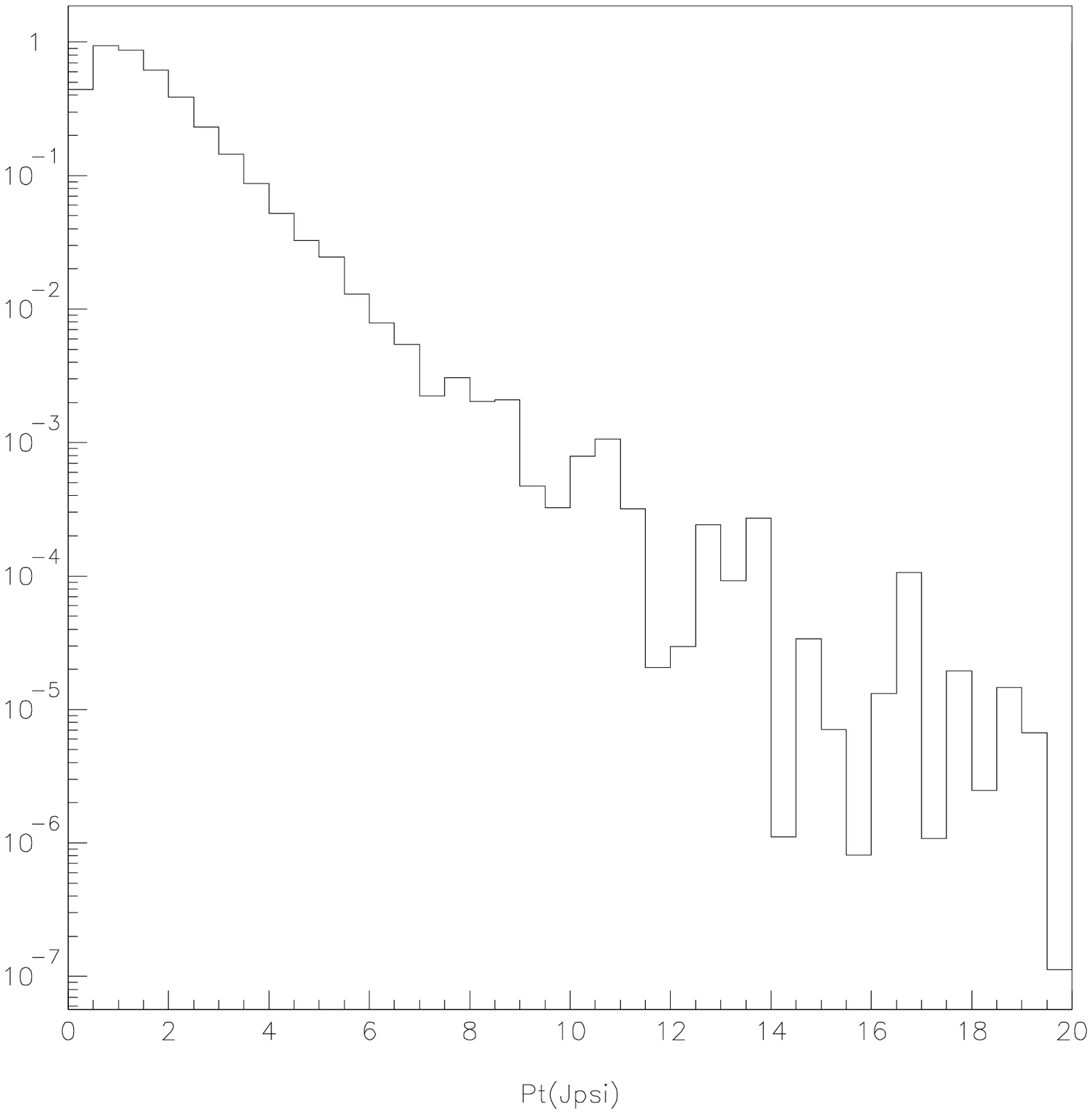}
\includegraphics[width=0.31\linewidth]{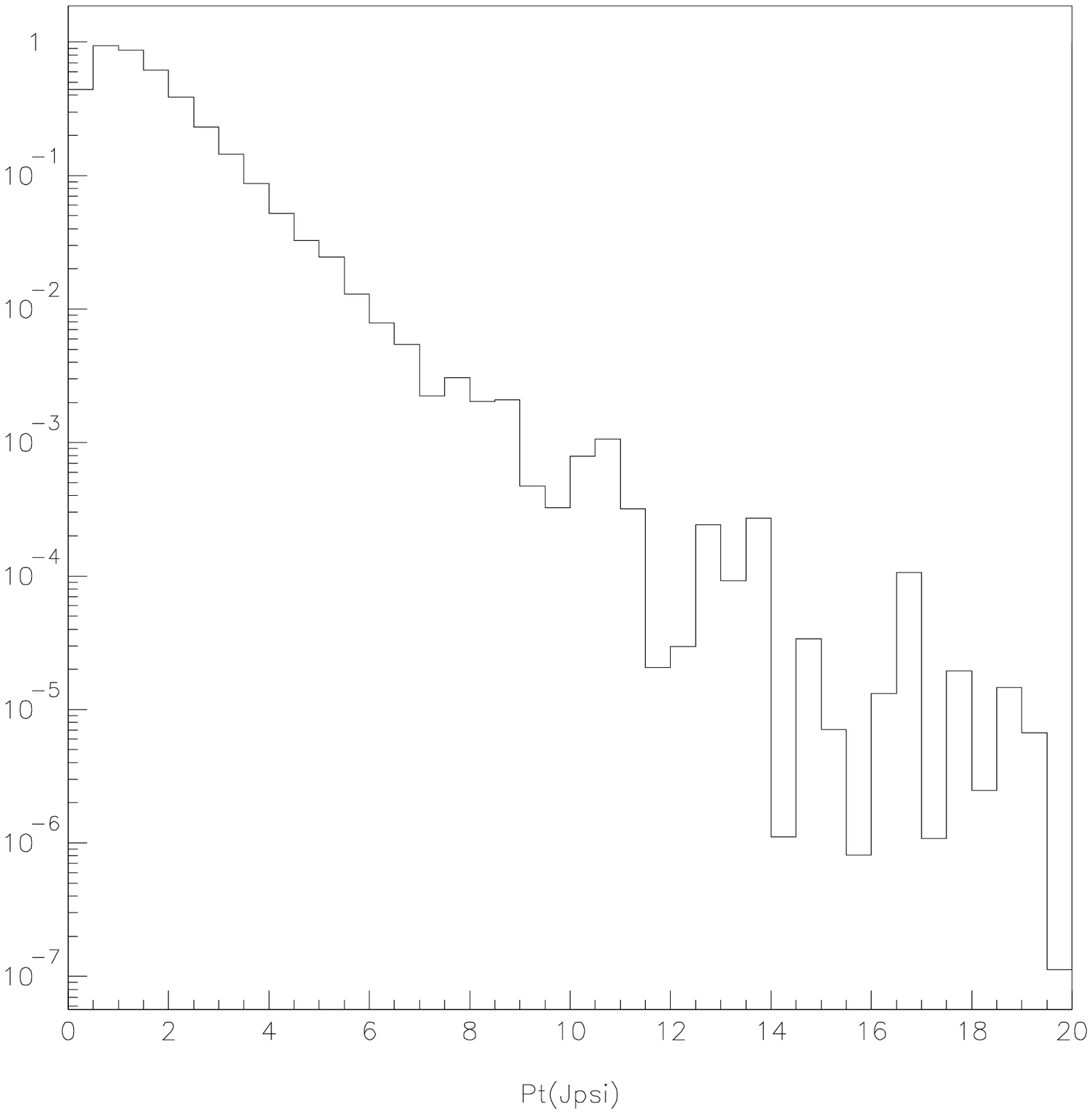}
\includegraphics[width=0.31\linewidth]{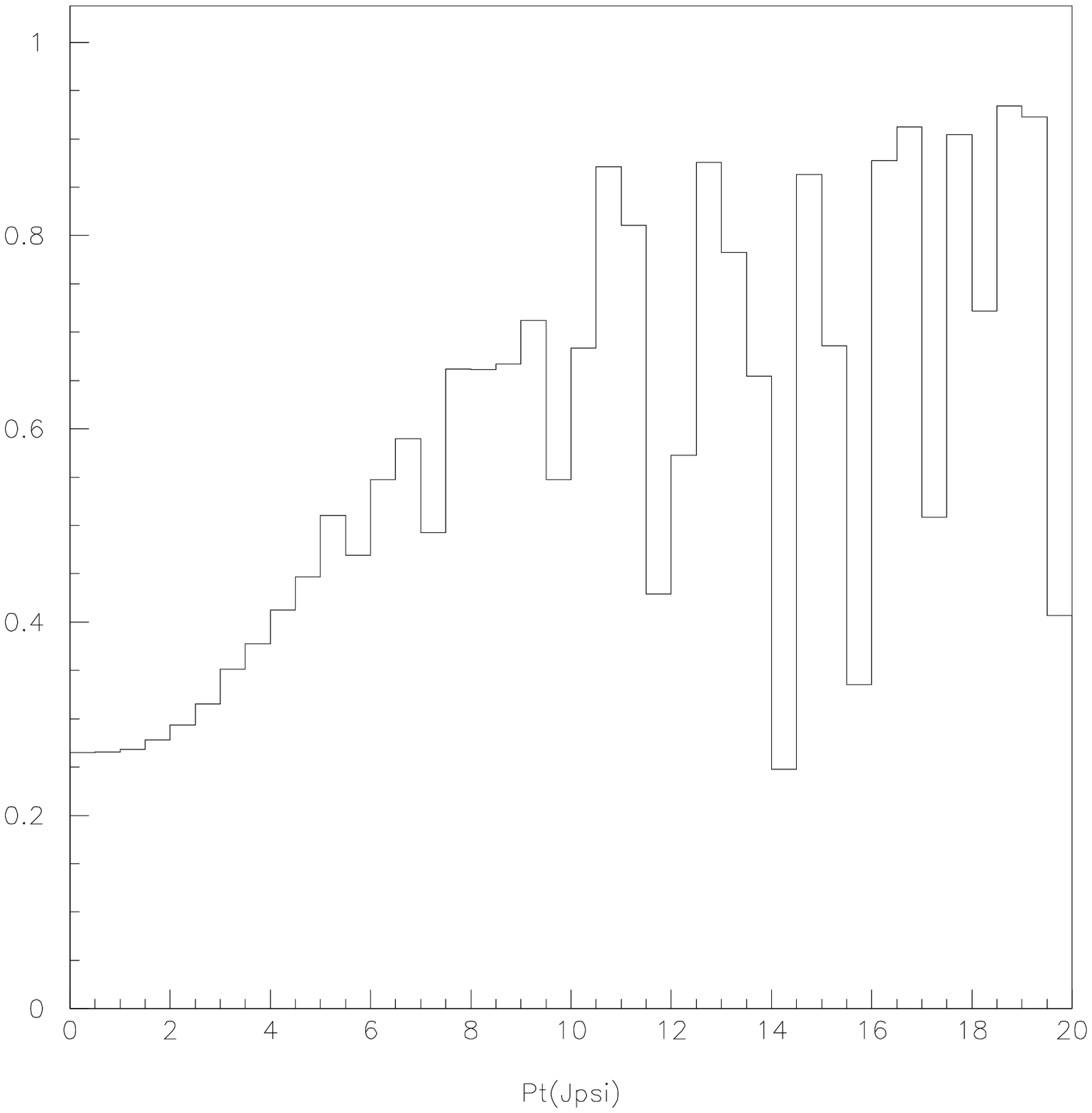}
\end{center}
\vspace*{-5mm}
\caption{{\it Differential cross sections for  inelastic $J/\Psi$ 
photoproduction with BFKL unintegrated gluon distributions
at THERA: (a) Inclusive $J/\Psi$ transverse momentum distribution,
(b) the same, but for $J/\Psi$ zero helicity states only,
(c) the fraction of $J/\Psi$ mesons in helicity zero state
(degree of spin alignement).}}
\end{figure*}

The role of the gluon virtuality may be seen in Fig. 3, where we show the results 
of  calculations  for \J \, photoproduction 
made with the BFKL  unintegrated gluon distribution.
The results correspond to THERA conditions, i.e. electron proton collisions
at the energy $\sqrt s =$ 1000 GeV, where no other cuts were applied except the
photoproduction limit $Q^2 < 1$ GeV$^2$ and the inelasticity requirement
$0.4 < z < 0.9$.

The effects of initial gluon off-shellness may be, best of all, seen in the transverse 
momentum spectra \cite{jbfkl}, because the gluon virtuality is proportional to its
transverse momentum: $m^2 = - k^2_{T}/(1 - x)$. In contrast with the conventional
(massless) parton model, the SHA shows that the fraction of \J \, mesons in the helicity
zero state increases with their transverse momentum $p_T$. A deviation from the parton
model behaviour becomes well pronounced already from $p_T > 3$ GeV at HERA energies
\cite{jbfkl}, and at $p_T > 6$ GeV the helicity zero polarization tends to be even
dominant (Fig. 3c).

Qualitatively, the difference between the model predictions referes to different
origins of the \J \, transverse momentum. In the case of conventional parton model
\J \, meson obtains its  transverse momentum from the hard photon gluon interaction,
while in the SHA there is also a large
 the contribution from the initial gluon transverse momentum.

The degree of spin alignement can be measured experimentally since 
the different polarization states of \J \, result in significantly different
angular distributions of the $J/\psi \to l^+l^-$ decay leptons:
\begin{equation}
d\Gamma_{h=\pm 1}/dcos{\Theta} = 1 - cos^2{\Theta}, \qquad
d\Gamma_{h=0}/dcos{\Theta} = 1 + cos^2{\Theta}
\end{equation}
Here $\Theta$ is the angle between the lepton and \J \,directions,
measured in the \J \, meson rest frame. Evidently the most informative
regions relate to $cos\Theta = \pm 1$.\\

\noindent {\bf 4.~Conclusions}\\

In the framework of semihard QCD approach, we obtained some predictions
for the cross sections of inclusive \dc  meson production at THERA conditions
using  different unintegrated gluon distributions driven by
the BFKL and CCFM evolution equations. 

We have considered the photoproduction of \dc mesons associated
 with two hadron jets and also \dc production in DIS at THERA conditions,
 which may be a sensitive indicator of the underlying
parton dynamics. The results of the simulations show
that theoretical results are very sensitive to BFKL type dynamics,
in particular, to the unintegrated gluon distribution in the proton.

We have considered also the effects of initial gluon  off-shellness
in  SHA for \J \, meson photoproduction at THERA energies.
Gluon virtuality connected with its transverse momentum is one of the 
inherent properties of noncollinear (BFKL) parton evolution theory.
Compared to traditional (collinear) parton model, gluons are characterized
by a different spin density matrix. The latter is found to affect the
polarization of \J \, mesons produced in $ep$ collisions via photon
gluon fusion subprocess. The effect is best pronounced  at large \J \,
transverse momenta and can be detected experimentally by measuring
 the $J/\psi \to l^+l^-$ decay lepton angular distributions.
We recommend the above process as a direct probe of the gluon
virtuality, which can eventually testify for the validity of 
BFKL gluon evolution.
  
Thus the experimental and theoretical investigations in the new
kinematic region of THERA collider will provide additional
tests of the semihard ($k_T$ factorization) approach and, in  particular, of the
"universality" of unintegrated gluon distribution.\\

\noindent {\bf Acknowledgements}\\

The work was supported by Royal Swedish Academy of Sciences.
~\\[2cm]
\end{document}